\documentclass[11pt]{article}
\usepackage{moriond,epsfig}

\def\Journal#1#2#3#4{{#1} {\bf #2}, #3 (#4)}

\def\NPB{{\em Nucl. Phys.} B}
\def\PLB{{\em Phys. Lett.}  B}
\def\PRL{\em Phys. Rev. Lett.}
\def\PRD{{\em Phys. Rev.} D}

\def\EPJ{{\em Eur. Phys. J.} C}
\def\JHEP{{\em JHEP}}

\def\D0{D\O}  \def\d0{D\O}

\def\mco{\multicolumn}

\def\be{\begin{equation}}
\def\ee{\end{equation}}
\def\bea{\begin{eqnarray}}
\def\eea{\end{eqnarray}}

\begin{document}
\vspace*{4cm}
\title{ELECTROWEAK PHYSICS:  $W$ AND $Z$ PRODUCTION AND STANDARD MODEL HIGGS SEARCH FROM THE FERMILAB TEVATRON}

\author{Abid Patwa$^{1,2}$}

\address{$^{1}$for the CDF and {D\O} Collaborations, \\
$^{2}$Department of Physics, Brookhaven National Laboratory,\\
Upton, New York 11973, USA}

\maketitle\abstracts{{\bf Abstract.} This paper reviews recent measurements by the {D\O} and CDF 
collaborations in $p\bar{p}$ collisions at the Fermilab Tevatron at $\sqrt{s} = 1.96$ TeV of $W$ and $Z$ 
boson production cross sections and asymmetries. Results from both experiments for the search of the 
Higgs boson predicted by the standard model in four different decay channels are also presented. 
}

\section{Introduction}

Production of $W$ and $Z$ bosons in high energy $p\bar{p}$ collisions has been precisely predicted 
by the standard model (SM). Experimental measurements of their production cross sections not only allow for 
stringent tests of the SM but also serve as a means to understand detector performance.   At hadron colliders, 
the hadronic decays of the $W$ and $Z$ tend to be overwhelmed by large QCD backgrounds and $W$ and $Z$ 
signatures can instead be efficiently identified through their leptonic decays.    This abundant source of 
high $p_{T}$ leptons also enables one to understand backgrounds for other important physics processes such 
as those in top, Higgs and SUSY decays.  
 
Moreover, the Higgs boson is the only particle predicted in the standard model that has not yet been 
discovered.   It is introduced in the SM in order to explain the mechanism of electroweak symmetry breaking.   
Experimental constraints provide considerable insight to its mass, which is a free parameter in the SM.  
Direct searches by the LEP experiments~\cite{lep} yield a lower limit of 114.4 GeV at 95\% C.L. on its mass while 
electroweak global fits give a 95\% C.L. upper bound at 219 GeV~\cite{lepwk}.   Present measurements of the $W$ and 
top mass further constrain the Higgs mass via radiative corrections and tend to favor a relatively 
light mass, placing it within the anticipated sensitivity range of present collider experiments.   Thus, 
the search for the Higgs is an integral part of the physics program at the Fermilab Tevatron experiments, CDF and {D\O}.  

This paper discusses measurements by CDF and {D\O} of the $W$ and $Z$ production cross sections 
in their leptonic decay channels and a recent result of the $W$ charge asymmetry by {D\O}.  Higgs 
searches within the SM in four different decay channels are also described.  By the winter of 2005-2006, 
both CDF and {D\O} have each recorded an integrated luminosity of over 1 fb$^{-1}$.   The results presented here 
use up to 0.4 fb$^{-1}$ of this dataset.

\section{$W$ and $Z$ Production Cross Sections}

Leptonic signatures of $W$ boson decays are characterized by an isolated, energetic lepton with substantial missing 
transverse energy whereas $Z$ boson decays contain two isolated, energetic leptons of opposite charge.    {D\O} and CDF 
each use a very similar selection technique in detecting electrons ($e$) decaying from the $W$ and $Z$.   Events must initially 
pass a single or di-electron trigger.   Offline, at least one electron with $p_{T} > 25$ GeV must lie within the central fiducial 
volume of the detector ($|\eta| < 1.0$) and, for the $W$ candidates, must have $\not\!\!E_T > 25$ GeV.   
For the 2$^{nd}$ electron from $Z$ candidates, {D\O} imposes very similar requirements as that of the 1$^{st}$ electron whereas 
CDF increases its electron acceptance to $|\eta| < 2.8$.    Using 72 pb$^{-1}$ integrated luminosity, CDF has selected 
37,584 $W$ and 4,242 $Z$ boson candidates.  {D\O} has analyzed a 177 pb$^{-1}$ data sample, which yields 116,569 
$W$ and 4,625 $Z$ boson candidates.    The difference in number of events is largely due to the different 
integrated luminosities between the experiments. 

Recently, CDF has also updated its $W\rightarrow e\nu$ cross-section measurement in the rapidity region 
1.1 $< |\eta| <$ 2.8.  Using the end-plug calorimeter and the ability to reconstruct 3D tracks at large $\eta$ with the SVXII 
silicon detector, the measurement studies $W$ reconstruction and properties in the forward region.  Here, with an 
integrated luminosity of 223 pb$^{-1}$, 48,144 $W$ boson candidates are selected.   

\begin{figure}
\begin{centering}
\psfig{figure=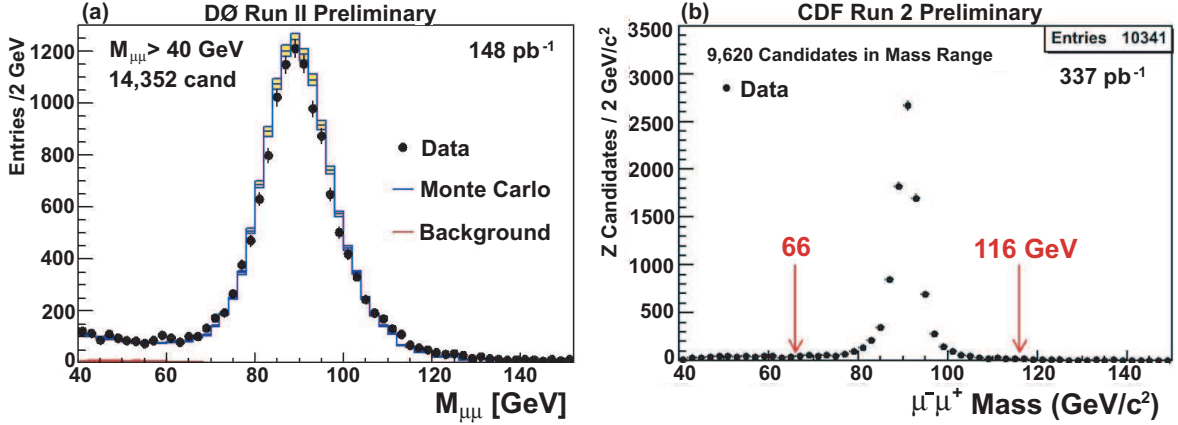,height=2.25in}
\caption{$Z\rightarrow\mu\mu$ invariant mass distribution for cross section measurements by (a) {D\O} and (b) CDF.
\label{fig:zmumu}}
\end{centering}
\end{figure}

For $W$ and $Z$ bosons decaying to muons ($\mu$), both CDF and {D\O} select events that pass either a single or di-muon 
trigger and offline, require an isolated track in their muon system matched to a track in the central tracking 
system.    Muons from $W$ candidates must have $p_{T} > 20$ GeV and $\not\!\!E_T > 20$ GeV.  For $Z$ candidates, CDF selects muons 
of $p_{T} > 20$ GeV, both within $|\eta| < 1.0$.   On the other hand, {D\O} requires each muon to satisfy 
$p_{T} > 15$ GeV and takes advantage of the wider acceptance of its muon chambers by including $\mu$'s up to $|\eta|=2.0$.  
CDF selects 57,109 $W$ boson events using a 194 pb$^{-1}$ data sample and 9,620 $Z$ boson events from a 337 
pb$^{-1}$ sample.   {D\O} selects 62,285 $W$ candidates from a 96 pb$^{-1}$ sample and 14,352 $Z$ candidates from a 
148 pb$^{-1}$ dataset.    Figure \ref{fig:zmumu} shows the invariant mass for $Z$ boson candidates after applying selection requirements.
The larger number of {D\O} events indicates the increased acceptance of its muon system while the narrower peak in CDF's 
invariant mass distribution of muon pairs demonstrates the higher CDF tracking resolution.

CDF has updated its production cross section measurement in the $Z\rightarrow\tau\tau$ channel with 350 pb$^{-1}$ of 
integrated luminosity.   Since $\tau$ leptons decay a short distance before reaching any detector into a) $e\nu_{e}\nu_{\tau}$, 
b) $\mu\nu_{\mu}\nu_{\tau}$, or c) hadrons $+ \nu_{\tau}$, effective $\tau$ identification algorithms must be used to 
discriminate between real taus (consisting of either charged leptons or narrow jets) with backgrounds dominated by 
jets produced by strong interaction processes.    CDF reconstructs $\tau$ candidates as narrow, isolated energy clusters 
in the calorimeter associated with charged tracks.   Next, $\pi^{\circ}$ information is added such that the invariant mass 
of the $\pi^{\circ}$ and the matched track-calorimeter cluster is consistent with the $\tau$ mass. 
  
\begin{figure}[h]
\begin{centering}
\psfig{figure=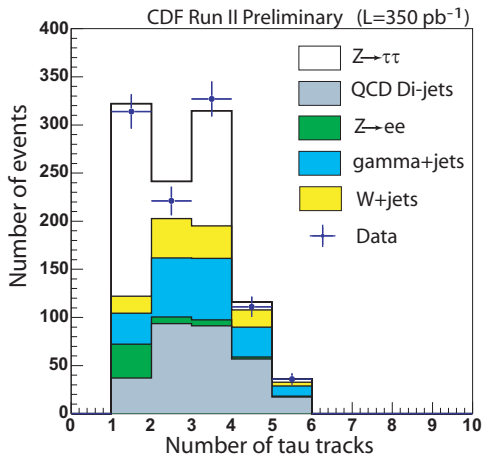,height=2.4in}
\caption{Track multiplicity distribution for a tau candidate in $Z\rightarrow\tau\tau$ events at CDF.
\label{fig:tau_trk}}
\end{centering}
\end{figure}

Reconstructing $Z\rightarrow\tau\tau$ candidates, CDF uses the channel where one $\tau$ decays into an electron and the 
other decays hadronically.   Event selections initially require one good quality isolated electron with $E_{T}^{e} > 10$ 
GeV and one hadronic tau candidate with $E_{T}^{\tau} > 15$ GeV, both in the central part of the 
detector, $|\eta_{e,\tau}| < 1.0$.   In order to suppress QCD and $W+$jet backgrounds, 
event topology cuts are imposed.   Figure \ref{fig:tau_trk} shows the track multiplicity distribution for observed hadronic taus and 
the expected backgrounds.  A total of 504 $Z\rightarrow\tau\tau$ candidate events pass the selection requirements 
with a 37\% QCD background estimate.

After all event selections, CDF and {D\O} calculate the product of $W$ and $Z$ production cross sections and 
branching ratios ($\sigma \times BR$) for each leptonic decay mode.   The results are summarized in Table 1.    
At present, the accuracies are limited primarily by systematic effects extending from 
a) lepton identification (for $e$, $\mu$: $\sim$1-2\%, for $\tau$: $\sim$3-4\%), b) use of PDF ($\sim$1-2\%), and 
c) background estimation (for $e$, $\mu$: $<$ 1\%, for $\tau$: $\sim$4-5\%).   In general, the $\sigma \times BR$ 
measurement uncertainties are dominated by the uncertainty on the luminosity measurement, which is 6.0\% (CDF) and 
6.5\% ({D\O}).    All measurements are in agreement with NNLO theoretical calculations~\cite{rham}.  Using the ratio 
$\sigma_{W} \times BR / \sigma_{Z} \times BR$, one can extract the total width of the $W$ boson, $\Gamma_{W}^{tot}$.  
Both the CDF and {D\O} result, also listed in Table 1, are in agreement with the SM value.  
Moreover, the CDF result for $\sigma_{W} \times BR(W\rightarrow e\nu)$ at forward rapidities is also 
consistent with their measurement in the central region.  The CDF $W\rightarrow e\nu$ and $Z\rightarrow ee$ measurements 
with the 72 pb$^{-1}$ data have been published~\cite{da}.  {D\O}'s  $\sigma_{Z} \times BR(Z\rightarrow\tau\tau)$ measurement, 
also given in Table 1, has been published~\cite{va1}.  

\begin{table}[t]
\caption{Summary of Tevatron $\sigma_{W,Z} \times BR$ and indirect $\Gamma^{tot}_{W}$ measurements.\label{tab:tev_cs}}
\vspace{0.4cm}
\begin{center}
\begin{tabular}{|c|c|c|}
\hline
&
\mco{2}{|c|}{$\sigma_{W,Z} \times BR \pm stat \pm sys \pm lum$ [in pb]} \\
&
\mco{2}{|c|}{($\int \mathcal{L} dt$)}\\
\cline{2-3}
                            & {D\O}                      & CDF \\ \hline
$ W\rightarrow\mu\nu_{\mu}$ & $ 2989\pm15\pm81\pm194  $  & $ 2786\pm12 ^{+65}_{-55} \pm 166 $       \\
                            &  (96 pb$^{-1}$)            &  (194 pb$^{-1}$)                         \\ \hline
$Z\rightarrow\mu\mu$        & $291\pm3.0\pm6.9\pm18.9$   &  $261.2\pm2.7 ^{+5.8}_{-6.1} \pm 15.1 $  \\
                            &  (148 pb$^{-1}$)           &  (337 pb$^{-1}$)                         \\ \hline
$W\rightarrow e \nu_{e}$    & $2865\pm8.3\pm76\pm186$    &  $2780\pm14\pm60\pm167$                  \\ 
                            &  (177 pb$^{-1}$)           &  (72 pb$^{-1}$, central)                 \\
\cline{3-3}
                            &                            &
\mco{1}{|c|}{$2815\pm13  ^{+94}_{-89}\pm169$} \\
                            &                            &
\mco{1}{|c|}{(223 pb$^{-1}$, end-plug)}        \\ \hline
$Z\rightarrow ee$           & $264.9\pm3.9\pm9.9\pm17.2$ & $255.8\pm3.9\pm5.5\pm15.4$                \\
                            & (177 pb$^{-1}$)            &  (72 pb$^{-1}$)                           \\ \hline
$\sigma_{W}/\sigma_{Z} \pm stat \pm sys$  &  $10.82\pm0.16\pm0.28$   &  $10.92\pm0.15\pm0.14$         \\
$\Rightarrow \Gamma^{tot}_{W}$            &  $\Rightarrow$ $2098\pm74$ MeV         &  $\Rightarrow$ $2079\pm41$ MeV               \\ 
\cline{2-3}
SM Theory ($\Gamma^{tot}_{W}$)            &
\mco{2}{|c|}{$2092.1\pm2.5$ MeV} \\ \hline

$W\rightarrow\tau\nu_{\tau}$&     $-$                    &  $2620\pm7.0\pm210\pm160$                 \\
                            &                            &  (72 pb$^{-1}$)                           \\ \hline
$Z\rightarrow\tau\tau$      & $237\pm15\pm18\pm15$       &  $265\pm20\pm21\pm15$                     \\
                            & (226 pb$^{-1}$)            &  (350 pb$^{-1}$)                          \\ \hline

\end{tabular}
\end{center}
\end{table}

\section{$W$ Production Charge Asymmetry}

Many electroweak measurements such as the $W$ mass measurement at hadron colliders depend on the theoretical 
calculation of the $W$ and $Z$ cross sections or their transverse momentum distributions.   At present, the 
precision of these quantities is limited by the uncertainties in the parton distribution functions (PDFs) used in these calculations.    
Studies of the $W$ production charge asymmetry probes PDFs and can help provide new PDF constraints.
	
In a $p\bar{p}$ collider, the primary production mode of $W^{+}$ bosons is $u+\bar{d} \rightarrow W^{+}$.   The $u$ quark 
carries more momentum than the $\bar{d}$ quark causing the $W^{+}$ boson to be boosted in the proton direction and 
similarly, the $W^{-}$ boson is boosted in the anti-proton direction.  This results in a forward-backward charge asymmetry 
in the rapidity distribution for positive and negative $W$ bosons.   Since the longitudinal component of the neutrino from the 
$W$ cannot be measured, the four momentum of the $W$ cannot be fully reconstructed.   
Alternatively, the observable quantity is the lepton ($\ell$) charge asymmetry, which is a convolution of the 
$W$ production asymmetry and the V-A decay of the $W$.   Since the latter is well understood, the charge asymmetry 
factors into $u$ and $d$ PDF, defined by

\begin{equation}
A(\eta_{\ell})=\frac{d \sigma (\ell^+)/d \eta - d \sigma (\ell^-)/d \eta}
{d \sigma (\ell^+)/d \eta + d \sigma (\ell^-)/d \eta}\simeq \frac{d(x)}{u(x)}
\label{eq:wasym}
\end{equation}

\begin{figure}
\begin{centering}
\psfig{figure=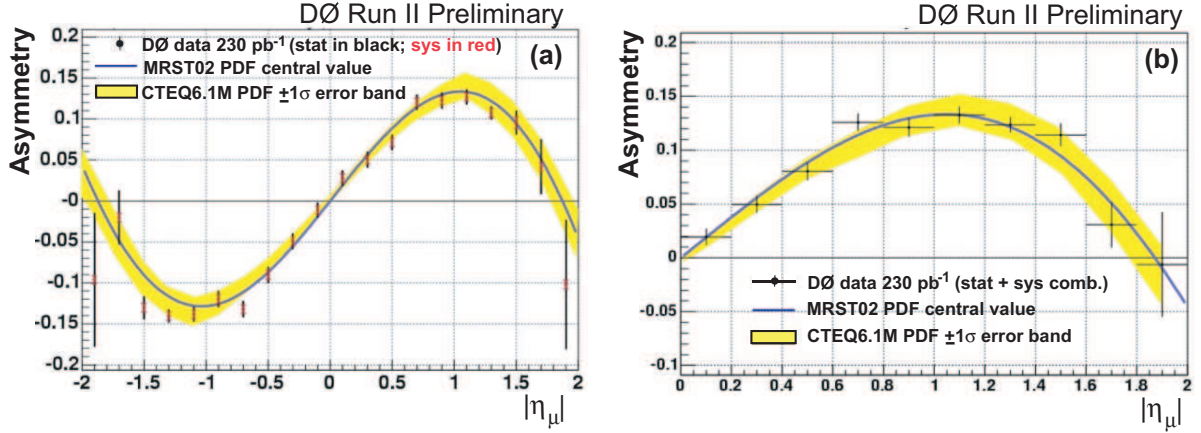,height=2.3in}
\caption{{D\O}'s measurement of (a) $W\rightarrow\mu\nu$ charge asymmetry and (b) with CP folding.  
Shown with the data (points) are the CTEQ6.1 PDF $\pm 1\sigma$ error band (yellow, shaded) and the central 
MRST02 PDF value (blue curve).  
\label{fig:wcharge}}
\end{centering}
\end{figure}

{D\O} has recently measured the $W$ charge asymmetry in the $W\rightarrow\mu\nu$ channel with a 230 pb$^{-1}$ sample 
of data. Since the phase space for the measurement at tree-level is limited by the range of $W$ boson rapidity that can be 
reconstructed, widest $\eta$ coverage is essential.  The large rapidity coverage of the {D\O} muon detectors together with 
forward muon triggers enables the asymmetry to be measured up to $|\eta|=2.0$.   Event selections require a single isolated 
muon with large missing transverse energy from the neutrino, similar to those used in the $W\rightarrow\mu\nu$ cross section 
measurement.  Proper charge identification is critical for the measurement; {D\O} imposes track quality requirements to 
improve the charge-id rate and measures the mis-id probability at 0.01\% for $|\eta|\sim 2.0$.   Figure \ref{fig:wcharge}a shows the measured 
muon charge asymmetry distribution corrected for background effects.   At large $\eta$, the measurement is statistically 
limited, and in order to improve the statistical uncertainties, the asymmetry is CP folded, as shown in Figure \ref{fig:wcharge}b.   
Also overlaid in Figure \ref{fig:wcharge} are the MRST02 PDF central value~\cite{mrst2} and the CTEQ6.1M PDF $\pm 1\sigma$ error bands~\cite{jp}.   A 
comparison between each indicates that {D\O}'s result can be used to help constrain future PDFs.
  
\section{SM Higgs Searches}

Tevatron searches for the Higgs ($H$) boson expected within the standard model rely on two basic search 
strategies.   For light mass Higgs ($M_{H} < 135$ GeV), the dominant production mode is via gluon fusion, where 
the Higgs subsequently decays into a $b\bar{b}$ jet final state.  Since this mode tends to be overwhelmed with 
multijet backgrounds, the search instead looks for the Higgs produced in association with either a $W$ or $Z$ boson.   
The final state consists of leptons from the $W$ or $Z$ in addition to two $b$-jets from the Higgs.   
Such analyses exploit efficient lepton identification and $b$-tagging techniques.   For higher mass Higgs 
($M_{H} > 135$ GeV), searches which give the most sensitivity concentrate on Higgs production via gluon fusion 
and its decay into two gauge bosons, primarily $WW$. The studies hinge around reconstruction of leptons from the $WW$.  

\subsection{$WH\rightarrow\ell\nu b\bar{b}$ Search Channel}

\begin{figure}[b]
\begin{centering}
\psfig{figure=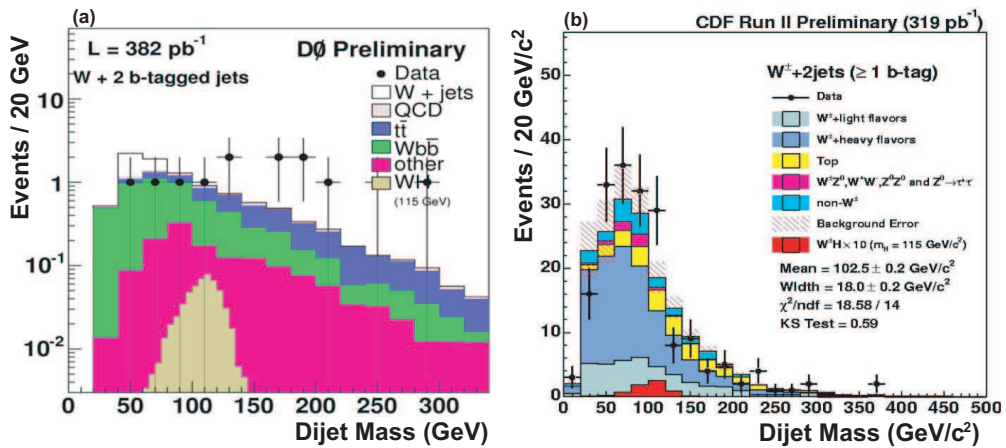,height=2.35in}
\caption{Dijet invariant mass distribution for Higgs search in $WH$ channel by (a) {D\O} and (b) CDF.
\label{fig:dijet}}
\end{centering}
\end{figure}

\begin{figure}
\begin{centering}
\psfig{figure=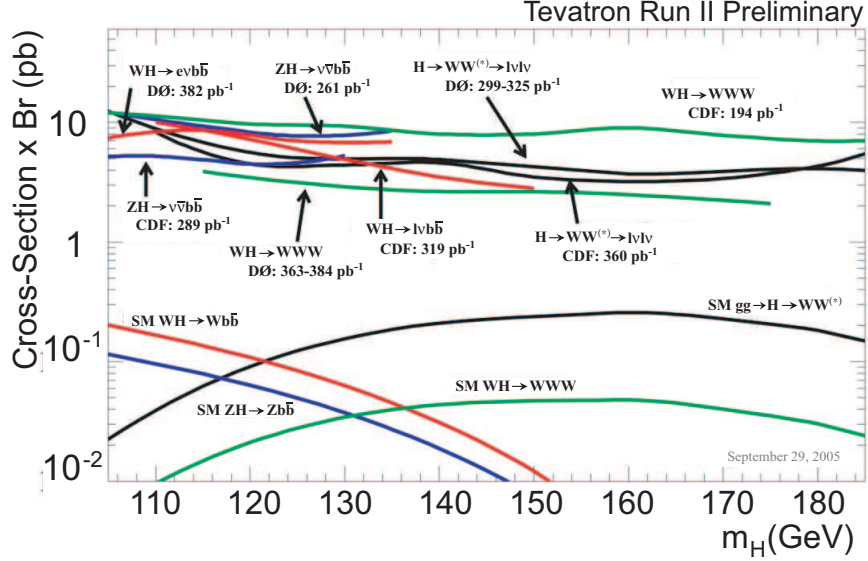,height=3.0in}
\caption{Summary of standard model Higgs boson searches at the Tevatron. Shown are the 95\% C.L. production cross section upper-limits as a 
function of Higgs masses measured by CDF and {D\O} across different Higgs channels.
\label{fig:higgs_cs}}
\end{centering}
\end{figure}

CDF and {D\O} use very similar methods to search for the associated production of the Higgs with a $W$ boson.    
{D\O} studies the channel $WH\rightarrow e \nu b\bar{b}$. Meanwhile, CDF not only studies the electron channel but 
also focuses on $WH\rightarrow\mu\nu b\bar{b}$.  The experimental signature requires a final state with one 
high $E_{T}$ lepton ({D\O}: $E_{T}^{e} > 20$ GeV, CDF: $E_{T}^{e} > 20$ GeV or $p_{T}^{\mu} > 20$ GeV), 
two $b$-jets, and significant missing transverse energy ({D\O}: $\not\!\!E_T > 25$ GeV, CDF:  $\not\!\!E_T > 20$ GeV).   
CDF selects events with at least one tagged $b$-jet while {D\O} imposes a tighter condition where the second 
$b$-jet must also be tagged.     The dominant backgrounds to $WH$ production are from $W+$heavy flavor, 
$t\bar{t}$, and single-top quark production.   Using a dataset of 382 pb$^{-1}$ at {D\O} and 319 pb$^{-1}$ at CDF, 
each experiment looks for a resonant mass peak in the dijet mass spectrum, as shown in Figure \ref{fig:dijet}.   
In order to limit the $Wb\bar{b}$ background contribution, the search is restricted to optimized 
$b\bar{b}$ invariant mass intervals, which vary with the Higgs masses studied.    
For $M_{H}=115$ GeV, {D\O} observes 4 events in a dijet mass window of 85 $< M_{b\bar{b}} <$ 135 GeV, 
and the expected SM background is 2.37 $\pm$ 0.59.    Similarly, CDF observes 14 events with a 14.62 $\pm$ 3.25 
expected background.  Across the different Higgs masses, both experiments find the signal to be consistent 
with the expected backgrounds and set 95\% C.L. upper limits on the cross section $\sigma_{WH} \times BR(H\rightarrow b\bar{b})$ 
at 7.6 to 6.9 pb ({D\O}) and 10 to 2.8 pb (CDF, single $b$-tag analysis) and 9.7 to 6.6 pb 
(CDF, double $b$-tag analysis).   The results are shown in Figure \ref{fig:higgs_cs}.

\subsection{$ZH\rightarrow\nu\bar{\nu} b\bar{b}$ Search Channel}

Because of the large $Z\rightarrow\nu\bar{\nu}$ and $H\rightarrow b\bar{b}$ branching ratios, one sensitive 
way to search for a light Higgs is its associated production with a $Z$ boson.   At low masses, 
the product of cross section and branching fraction is expected to be on the order of 0.01 pb and comparable to that of 
$WH\rightarrow\ell\nu b\bar{b}$.   Since the two $b$-jets from the Higgs boson decay are boosted along the Higgs 
momentum direction, the final state contains a distinct signature of acoplanar jets in contrast to 
typical back-to-back QCD dijets.   The main backgrounds are from $W/Z+$jets, electroweak diboson ($WZ$ and $ZZ$) 
production, and $t\bar{t}$ processes where the lepton or jets escape.  Multijets, where the jets are mismeasured and/or 
misidentified, dominate as instrumental backgrounds and contribute to the missing transverse energy measurement.

The {D\O} event selections that help separate the signal from background require two $b$-tagged jets with $p_{T} > 20$ 
GeV,  $\not\!\!E_T > 25$ GeV, no back-to-back event topology, and no isolated tracks in the event.   The last three conditions 
suppress QCD multijet and $W/Z+$jet background events.   Additional track and asymmetry cuts developed using variables 
such as the vector sum of the $p_{T}$ of all tracks, $\not\!\!E_T$, and $H_{T}$ (the scalar sum of jet $p_{T}$) 
reduce the instrumental backgrounds.
	
Event selections by CDF consist of two $b$-jets, each with $E_{T} > 25$ GeV, and at most a third soft jet, which can 
be radiated off by one of the b-jets.   While {D\O} requires both jets to be tagged, CDF selects events with at least one 
$b$-tagged jet.   Further, in order to limit the effect of jets or leptons being mismeasured, a $\not\!\!E_T > 70$ GeV condition 
is imposed.   The data is subsequently divided into two control regions to understand the backgrounds:  I) events 
with no high-$p_{T}$ isolated leptons and azimuthal angular separation between the second leading jet and $\not\!\!E_T$, 
$\varphi$(2$^{nd}$ jet, $\not\!\!E_T$)$<$0.4 and II) events with at least one lepton or isolated track and 
$\varphi$(2$^{nd}$ jet, $\not\!\!E_T$)$>$0.4 .   Region I is dominated by QCD multijet events and Region II contains top and 
electroweak backgrounds.    After comparing the data from these control regions to the simulated SM backgrounds, 
selections are optimized using variables defined by the $E_{T}$ of the leading jet, $H_{T}$, and the angular separation 
between the leading jet and $\not\!\!E_T$.
   
Both {D\O} (using 261 pb$^{-1}$ dataset) and CDF (using 289 pb$^{-1}$ dataset) search for a peak in the dijet 
invariant mass distribution.   Since no significant excess is observed over expected backgrounds, 95\% C.L. 
cross section upper limits are established at $\sigma_{ZH} \times BR(H\rightarrow b\bar{b})$  of 7.7$-$12.2 pb for Higgs masses 
between 105 and 135 GeV ({D\O}) and 4.5$-$5.45 pb between 90 and 135 GeV Higgs masses (CDF).   
These results also appear in Figure 5.  

\subsection{$H\rightarrow WW^{(*)}$ Search Channel}

Standard model Higgs search in $H\rightarrow WW^{(*)}$, which subsequently decays into three final states: $e^{+}e^{-}$, 
$e^{\pm}\mu^{\mp}$, and $\mu^{+}\mu^{-}$ is the dominant production mechanism for higher mass Higgs.   
Such events are triggered by the presence of isolated, oppositely charged single or di-leptons.  However, 
a Higgs search in this channel contains an exhaustive list of backgrounds: electroweak diboson ($WW$ and $WZ$) production, 
$Z/\gamma^{*}$, $ZZ$, where one or two leptons are respectively mismeasured, multijets and $W+$jets, where the 
jets are misidentified as leptons, and $t\bar{t}$ production.

Within this channel, both CDF and {D\O} use a very similar search strategy that is based on the $WW$ 
decay topology.  Due to the scalar nature of the Higgs, the two charged leptons from the $W$ are emitted in-parallel 
with small azimuthal angular separation ($\Delta\varphi_{ll}<2.0$) between them, and in contrast to the 
standard back-to-back multijet background. Moreover, from $W$-helicity conservation, the lepton system and the neutrinos 
are emitted mostly back-to-back, which allow the dilepton invariant mass to be constrained to $M_{H}/2$.    
Selections based on two oppositely-charged high $p_{T}$ leptons, large $\not\!\!E_T$ from the $\nu$'s, $\Delta\varphi_{ll}$, 
and dilepton invariant mass become an effective method in discriminating a Higgs signal from backgrounds.   
Once the selections have been optimized, the number of events observed by each experiment is consistent with those 
expected from backgrounds.  As summarized in Figure 5, 95\% C.L. production limits are therefore extracted for different Higgs 
masses.  The limits are from a combination of all three decay channels using integrated luminosities 
of CDF: 360 pb$^{-1}$ and {D\O}: 325$\pm$21 pb$^{-1}$ ($e^{+}e^{-}$), 318$\pm$21 pb$^{-1}$ 
($e^{\pm}\mu^{\mp}$), and 299$\pm$19 pb$^{-1}$ ($\mu^{+}\mu^{-}$).  The {D\O} result has been published~\cite{va2}.

\subsection{$WH\rightarrow WWW^{(*)}$ Search Channel}

The search for associated Higgs production via the channel 
$WH\rightarrow WWW^{(*)}\rightarrow\ell^{\pm}\nu\ell^{'\pm}\nu q\bar{q}$  is more promising 
than direct Higgs production, $H\rightarrow WW^{(*)}$, as it requires like-sign leptons and 
therefore, avoids large SM backgrounds from electroweak diboson and $t\bar{t}$ production, 
which contain oppositely charged leptons.    Instead, the primary physics background is from 
$WZ\rightarrow\ell\nu\ell\ell$, where one of the leptons from the $Z$ is lost.   Instrumental 
backgrounds that contribute non-negligibly to this mode are 
a) dominated by mismeasuring the charge of one of the leptons in $Z/\gamma^{*}\rightarrow\ell\ell$
decays (often referred to as ``charge flips'' at {D\O}) and b) QCD processes, which contain 
semileptonic heavy flavor decays, punch-through hadrons misidentified as muons, or $\gamma\rightarrow e$ conversions.
   
With 363$-$384 pb$^{-1}$ dataset, {D\O} selects an event containing like-sign $ee$, $e\mu$, and $\mu\mu$ candidates with 
$p_{T} > 15$ GeV per lepton and  $\not\!\!E_T >20$ GeV.   Track quality cuts are imposed to reduce the charge flip 
probability. The background composition differs among the three channels.  Because of the improved $\not\!\!E_T$
measurement in the $ee$ channel, the $\not\!\!E_T$ cut is very effective in reducing QCD and 
charge flips and thus, the $ee$ channel is dominated by the $WZ$ physics background.   In contrast, charge flips 
dominate the $\mu\mu$ channel (where $Z/\gamma^{*}\rightarrow\ell\ell$ production is significant) and QCD 
contribute largely to the $e\mu$ channel. Nonetheless, {D\O} observes 1 $ee$ event, 3 $e\mu$ events, and 
2 $\mu\mu$ events, all which are in agreement with the expected backgrounds.

CDF's present analysis based on an integrated luminosity of 194 pb$^{-1}$ stems from simple techniques that 
require isolated like-sign leptons but do not use explicit selections on the signal such as $\not\!\!E_T$ and other 
topological cuts. Instead, the data is divided into a signal and a series of control regions determined by the 
magnitude and vector sum of the first and second leading lepton $p_{T}$.   No events are observed in the signal 
region, while the total background is expected to be 0.95$\pm$0.80.    

Again, Figure 5 shows the 95\% C.L. upper limits on the cross section as a function of different Higgs masses.   
Since no signal-specific cuts are introduced in the analysis, the present CDF result is conservative for the Higgs 
search.   However, once further optimizations are done, sensitivity in this production mode seems very promising. 

\section{Conclusion}

Using a fraction of the 1 fb$^{-1}$ integrated luminosity, CDF and {D\O} have measured the product of cross section 
and branching ratios in the different leptonic channels of $W$ and $Z$ boson production and all results 
are in agreement with standard model expectations.   The measurements also lay the foundation for understanding 
not only the detector response but also backgrounds for other important physics processes.

Moreover, both CDF and {D\O} have searched for the Higgs boson predicted in the standard model across a 
comprehensive set of search channels.  In the absence of signal, each experiment has established 95\% C.L. cross section 
upper limits.   However, the analyses are presently trying to reach the sensitivity outlined by the 2003 Higgs Sensitivity 
Studies~\cite{hss}.   In particular, the current results from the Tevatron indicate the expected sensitivity has not been 
reached typically by a factor of 2$-$3.  Several approaches such as optimizing analyses techniques, adding or 
combining search channels, and combining results from each experiment will help bridge the gap between 
the current limits and those predicted by the SM.   Indeed, studies implementing some of these methods are 
already underway as are Higgs searches using the full 1 fb$^{-1}$ of collected data.   The Tevatron experiments look forward 
to the prospects for a light mass Higgs discovery with 4$-$8 fb$^{-1}$ of integrated luminosity.

\section*{Acknowledgments}
The author appreciates and wishes to thank the CDF and {D\O} collaborations for useful discussions and 
providing the results presented here.

\end{document}